# Measurement of BAS-TR imaging plate response to energetic aluminum ions

**J. Won,**[1,2] **J. Song,**[1,2] **S. Palaniyappan,**[3] **D. C. Gautier,**[3] **W. Jeong,**[1] **and J. C. Fernández,**[3] **W. Bang**[1,2,a]

[1]*Department of Physics and Photon Science, GIST, Gwangju 61005, South Korea*
[2]*Center for Relativistic Laser Science, Institute for Basic Science, Gwangju 61005, South Korea*
[3]*Los Alamos National Laboratory, Los Alamos, New Mexico 87545, USA*

**Abstract**
We measured the response of BAS-TR imaging plate (IP) to energetic aluminum ions in the 0 to 222 MeV energy range, and compared it with predictions from a Monte Carlo simulation code using two different IP models. Energetic aluminum ions were produced with an intense laser pulse, and the response was evaluated from cross-calibration between CR-39 track detector and IP energy spectrometer. For the first time, we obtained the response function of the BAS-TR IP for aluminum ions in the energy range from 0 to 222 MeV. Notably the IP sensitivity in the exponential model is nearly constant from 36 MeV to 160 MeV.

[a]Author to whom correspondence should be addressed. Electronic mail: wbang@gist.ac.kr

**Introduction**

The imaging plate (IP) is a film-like image sensor that records a radiation flux on a thin sheet called a phosphor layer. IPs are sensitive to energetic charged particles, X-rays, and gamma rays.[1-4] IPs have been widely used in physics and in medicine since the IPs were developed by Fuji film Co. in the early 1980s due to many advantages. Compared with other detectors, IPs have several advantages: (1) immunity to electromagnetic pulse (EMP), (2) high dynamic range (4-5 orders of magnitude), (3) high spatial resolution (resolving to as low as 10 µm), and (4) reusability (signal of IP can be erased with white light).[1, 5-7] BAS (Biological Analysis System)[8] IP types are commonly used for radiation detection. Specifically, BAS-MS, SR and TR, were primarily designed to optimize high sensitivity, high resolution, and detection of beta particles from tritium, respectively.[9-13] BAS-TR can detect beta particles produced by tritium as well as radiations that can be measured by BAS-MS and SR. BAS IPs typically consist of 3 or 4 layers having various thicknesses: a protective layer, a phosphor layer, a support layer, and a magnetic layer. The structure of BAS-TR IP, which has no protective layer, is shown in Fig. 1. The lack of a protective layer makes it particularly well suited to measure heavy ions because of their short range within matter. Incoming ions deposit kinetic energy in the phosphor layer ($BaFBr_{0.85}I_{0.15}:Eu^{2+}$, density 2.85 g/cm$^3$)[14]. Beneath the support layer, the magnetic layer allows magnetic attachment inside the scanner.[15]

When an IP is exposed to radiation, the electrons of $Eu^{2+}$ in the phosphor layer are ionized and trapped in FBr or FI sites forming metastable states. The lifetimes of the metastable states range from 10 minutes to a few days. When a scanner irradiates the phosphor layer with 2 eV photons from a laser diode, the electrons in the metastable state are re-excited and recombine with $Eu^{3+}$ and emit 3 eV photons. That emission is known as the photostimulated luminescence (PSL). A photomultiplier tube (PMT) equipped with a scanner converts the PSL to electrical signals and amplifies them. Once scanned, white light can erase any information remaining in the IP, de-exciting the electrons in metastable states to ground states, thus making the IP reusable.

In order to use an IP as a quantitative radiation detector, it is necessary to calibrate the PSL relative to the spectral intensity of radiation. This must be done for each type of radiation and IP and scanner combination, because each combination has in general a different calibration. [5, 7, 16-19] For ions, the spectral response function of an IP is measured in units of PSL per incident ion of a given energy.

Researchers have calibrated the response of IPs to various radiation types for a wide energy range. In particular, the IP responses to electrons, protons, and photons (X-ray or γ-ray) have been extensively studied previously. For protons, IP response calibrations have been reported by Mančić *et al.*[20] for BAS-TR within an incident energy range of 0.5-20 MeV, Choi *et al.*[21] (0.5–1.6 MeV, BAS-TR), Freeman *et al.*[22] (0.3–3.2 MeV, BAS-TR), and Bonnet *et al.*[17] (0.6–3.2 MeV, BAS-MS, SR, and TR). In 2005, Tanaka *et al.*[23] investigated the response of BAS-SR to electrons with 11.5, 30, and 100 MeV using a LINAC. Chen *et al.*[24] examined the response of BAS-SR to electrons for the range of 0.1 to 4 MeV, and Boutoux *et al.*[7] studied the IP response of five BAS types to 5–18 MeV electrons. Nakanii *et al.*[25] reported on the response of BAS-SR to 1 GeV electrons. BAS calibration for energetic heavy ion beams, however, is more difficult because conventional accelerators are not well suited for producing heavy ion beams with sufficient particle fluence for IP calibration at a range of energies in a reasonably short time. For this reason, the published studies of IP response to ions are limited to few types of heavier elements such as deuterium,[22, 26] helium,[19, 22], and carbon[6].

Since Hidding *et al.*[27] proposed a linear model predicting the response of IP, some researchers have calculated the IP response by calculating stopping powers using Monte-Carlo simulation codes. Bonnet *et al.* designed an exponential model and calibrated the IP response to protons, electrons, photons, and $^4$He ions, using Geant4.[17, 19, 28] Recently, Rabhi *et al.* reported on the calibration of BAS-MS, SR, and TR for 1–200 MeV protons[15] and for 40–180

MeV electrons[29] using Geant4. Singh *et al.* used FLUKA to calibrate the responses of BAS-MS and SR to 150 keV–1.75 MeV electrons.

In this paper, we report on the IP response to aluminum (Al) ions within the 0–222 MeV range for the first time. We have used an Al ion beam driven by an intense laser pulse as the ion source, and detected these ions using BAS-TR IPs for the calibration. We compare both the exponential model and the linear model with our experimental results for Al ions. In our study, we have used the Monte Carlo simulation code SRIM[30] to calculate the stopping powers of a BAS-TR IP for Al ions because SRIM is known to describe the available experimental data best.[31, 32] We show that the response function calculations using stopping power from SRIM code agree very well with our experimental measurements of the response function for Al ions in the 0–222 MeV energy range.

**Experimental setup**

The experiments were performed on the Trident laser facility at Los Alamos National Laboratory (LANL).[33-37] Figure 2 shows the schematic layout of the experimental setup. 80 J, 650 fs, 1054 nm laser pulses were focused using an f/3 off-axis parabola, and irradiated 110 nm thick aluminum foils with a peak laser intensity of about $2\times10^{20}$ W/cm$^2$.[33] The laser-driven Al ion beams diverged with a 20° cone half-angle.[35] The ions fly into a Thomson Parabola Spectrometer (TPS) which measures a spectrum separately for each individual charge to mass ($Z/A$) ratio, in our case Al$^{11+}$. The TPS symmetry axis is aligned with the ion propagation direction, and the ion flux into the TPS is limited by a pinhole aperture along that axis. Over a portion of the ion flight within the TPS, strong electric ($E$) and magnetic ($B$) fields parallel to each other and normal the symmetry axis deflects ions depending on their $Z/A$ and kinetic energy ($E_{\text{ion}}$). After a drift distance within the TPS, they arrive at the detector plane laid normal to the axis. The $Z/A$ and $E_{\text{ion}}$ is given by their location on this plane.[33, 38] Specifically, the TPS disperses a given *Z/A* on the detector plane along a narrow (as defined by the pinhole) parabolic curve in the $\hat{x}$ (horizontal) and $\hat{y}$ (vertical) directions originating at the intersection with the symmetry axis according to $E_{ion}$. The origin corresponds to infinite $E_{ion}$, and where any neutral particle would be recorded. The IP used as a TPS detector was covered with an 18 μm thick Al filter in order to reduce background noise owing to ambient light, low energy protons, electrons, and X-rays.[33] Al ions also lose kinetic energy on the Al filter, and only ions with kinetic energy greater than 50 MeV did not range out within the filter and reached the IP surface. For counting the absolute number of aluminum ions, strip-shaped CR-39 track detectors with a width of a few mm were placed on the IP surface.[33]

It is known that the IP response does not depend on the charge state of the incident ions.[6, 22, 39, 40] This is because the incident ions quickly arrive at an equilibrium charge state as soon as they enter the target surface. This characteristic is assumed by Freeman *et al*.[22], and is confirmed by Doria *et al*.[6] in an experiment using multiply charged carbon ions. Therefore, we are not required to specify the charge state of Al ions incident on the BAS-TR IP in our SRIM calculations.

**Measurement of PSL/ion**

After exposing the calibration region of an IP to energetic aluminum ions, PSL is measured experimentally using a scanner. In this experiment, we scanned each IP 5 minutes after exposing it to an ion beam. The BAS-TR IP was scanned by a commercial scanner (Fuji FLA-7000). The input light in the scanner is converted to electronic signals, which are stored in a PC as quantum level (QL) pixel-resolved image data. Since QL encodes a logarithmic response, a conversion is required to extract the linear PSL data. The conversion formula from a QL value to a PSL value is[6]

$$\text{PSL} = \left(\frac{R}{100}\right)^2 \times \frac{4000}{S} \times 10^{L \times (\frac{QL}{G} - \frac{1}{2})}, \qquad (1)$$

where $R = 25$ μm is the resolution of the scanner, $S$ is the scanner sensitivity varying from 1000 to 10000, and $L = 5$ is called the latitude. $G$ is the gradation, 65535 for 16 bit.

The PSL/ion value was obtained from cross-calibration between the amount of PSL scanned in the $Al^{11+}$ trajectory adjacent to a CR-39 strip and the number of ions counted in the pits of the strip. This calibration method assumes that there is no sharp discontinuity in the areal density of $Al^{11+}$ ions on the track nearby the edges of the CR-39 bars. These calibration regions are illustrated in Fig. 3, which shows the locations of the CR-39 strips. $E_{ion}$ is readily obtained from the TPS data based on the analytic expressions for the ion deflection at the detector plane. The x-direction deflection (due to the known $E$) is used to obtain $E_{ion}/q$, where $q$ is the ion charge. That value is plugged into the equation for the y-direction deflection (due to the known $B$) to obtain $Z/A$. Since we know $A$ (Al), we obtain $q$ and $E_{ion}$. Thus PSL/ion value for $Al^{11+}$ ions can be obtained for the $E_{ion}$ of $Al^{+11}$ ions corresponding to each edge of the CR-39 strip along the track.

**Fading effect**

As time elapses after activation, some of the electrons in the metastable states spontaneously decay to the $Eu^{3+}$ state and emit PSL. This phenomenon is called the fading effect, and the resulting PSL loss when the IP is scanned subsequently should be taken into account in the analysis. Several studies have measured fading curves for electrons[16, 23], protons[17], X-rays[5, 7, 18, 26], and γ-rays[17]. Although the radiation sources used in fading measurements are different in each experiment, fading curves are not very sensitive to the type and energy of radiation. Bonnet *et al.* finds that fading signals are nearly independent of the radiation type with less than 10% differences between photons and protons.[17] Ohuchi *et al.* also reports that the fading effect is similar for electrons and for protons regardless of their kinetic energies.[41] The known parameters contributing to fading effect are ambient temperature and scanner type.[7] Although the fading effect becomes bigger as the ambient temperature increases, its change can be considered negligible for a small temperature fluctuation.[7]

Zeil *el al.* showed different fading behaviors between his data measured by BAS-1800II and the data of Tanaka *et al.*[23] measured by FDL-5000, and the discrepancy was about 20%.[16] Ohuchi *et al.* also compared BAS-1000 and BAS-5000 and observed that fading of BAS-5000 is larger than that of BAS-1000.[41] Therefore, we need to apply a fading model benchmarked using the same IP. Bonnet *et al.*[17] and Boutoux *et al.*[7] used the same FLA-7000 scanner like in our measurements, and Fig. 4 shows measured fading decays for 200 minutes after irradiation. They used two exponential functions to fit their data. Figure 4 shows a 10% decrease within 5 minutes. We adopt the fading function of Boutoux *et al.* as shown below.[7]

$$f(t) = 0.535\, e^{-t/\tau_1} + 0.465\, e^{-t/\tau_2}, \qquad (2)$$

where $\tau_1 = 23.812$ (*min*), $\tau_2 = 3837.2$ (*min*). In our experiment, fading was expected to be about 10%, and thus the experimentally measured PSL values have been scaled upwards to the values that would have been measured at time zero.

**Calculation of PSL/ion from SRIM data**

There are two models commonly used to predict the amount of PSL from a given radiation. Hidding *et al.*[27] proposes a linear model assuming that the yield of PSL is proportional to the total deposited energy in the sensitive layer of an IP.

$$R(E_{Ion}) = \alpha E_{dep}(E_{Ion}), \qquad (3)$$

where $R(E_{Ion})$ is the IP response for ions with incident kinetic energy of $E_{Ion}$ and $\alpha$ is the IP sensitivity. The IP sensitivity varies depending on the type of radiation and IPs. It also depends on the waiting time before scanning because of the fading loss. The total deposited energy, $E_{dep}(E_{Ion})$, is obtained from the integral of the ion-stopping power *S(z)* at depth *z*. In the phosphoric layer of the IP,

$$E_{dep} = \int_0^W S(z) dz, \qquad (4)$$

where $S(z) = -dE_{ion}/dz$ and *W* is the thickness of the layer.

The incident ion loses its kinetic energy to electrons in the target by ion-electron collisions resulting from Coulomb interaction and to target nuclei by ion-nucleon collisions (called recoil process). In our SRIM calculations, the predominant process is Coulomb interaction, and the recoil energy contributes only ~$10^{-4}$ to the entire collision process. The cross-section for ion-electron interaction is inversely proportional to the square of the approaching speed of the ion. In general, slow ions lose larger amount of energy to target electrons than fast ions because they spend more time interacting with electrons.

Bonnet *et al.*[17] proposed a model for the deposited energy that accounts for the optical thickness of the IP to the PSL radiation by weighting the stopping power by an exponential decay term, *i.e.*,

$$E_{dep} = \int_0^W S(z) e^{-\frac{z}{L}} dz, \qquad (5)$$

where *L* is an absorption length. Based on the experimental measurements by Bonnet *et al.*, *L* = 44 ± 4 µm for BAS-TR IPs.[19] The absorption length can be understood as the mean free path of PSL photons in the phosphor layer.[17] A large *L* implies that the PSL traverses through the phosphor layer easily without being absorbed within it,[19] and thus a negligible correction. However, *L* = 44 µm indicates that a significant amount of PSL is absorbed since the thickness of BAS-TR IP is only 50 µm. Equation (2) can also be interpreted as a special case when the absorption length is infinity and there is no absorption.[17] We refer IP models proposed by Hidding *et al*. and Bonnet *et al*. as the linear model and the exponential model, respectively, in the following analysis.

To obtain the stopping power of Al beams, we use the Monte Carlo simulation code SRIM which calculates the stopping and range of ions in matter using a quantum mechanical treatment of ion-atom collision.[42] Each simulation is performed with 10,000 trials and calculated stopping power and other values are averaged over the 10,000 ion incidences. We calculate $E_{dep}$ using two different methods (1) by using the averaged *S(z)* obtained directly from SRIM, and (2) by calculating the absorbed energy and $e^{-\frac{z}{L}}$ at each step from individual trial then averaging $E_{dep}$ over 10,000 trials. SRIM divides the target depth into 100 steps in the first method, and about 700 uniform steps in the second method, which determines the simulation resolution. Although the second method has 7 times higher target depth resolution in computing the total deposited energy in each thin layer, the discrepancy in the deposited energy calculations by each method is found to be only 0.01–0.1 percent. In the linear model, we have calculated *S(z)* using the first method for simplicity. In the exponential model, we have used the second method to calculate *S(z)* since more steps can potentially reduce errors involved in calculating the weighting factor $e^{-\frac{z}{L}}$. For each IP model, the total deposited energy is calculated as follows

$$E_{dep} = \sum_{i=1}^{l} [\frac{S(z_i)+S(z_{i+1})}{2}](z_{i+1} - z_i) \quad \text{(Linear model)} \quad (6)$$

$$E_{dep} = \sum_{i=1}^{l}(-\frac{dE}{dz})e^{-\frac{1}{L}(\frac{z_{i+1}+z_i}{2})}dz$$

$$= \sum_{i=1}^{l} -(E_{i+1} - E_i)e^{-\frac{1}{L}(\frac{z_{i+1}+z_i}{2})}, \quad \text{(Exponential model)} \quad (7)$$

where $z_i$ is the target depth (the distance that an Al ion passes through the target), $S(z_i)$ is the stopping power of target at depth $z_i$, $E_i$ is the kinetic energy of the Al ion at the $z_i$, and $l$ is the total number of steps ($l = 100$ or $l = 700$). In the linear model, the area under the curve of stopping power as a function of target depth represents the deposited energy. We took the average of two stopping power values in each interval. We also estimate the target depth of each thin layer as $\frac{z_{i+1}+z_i}{2}$ in the exponential model.

The average stopping powers of phosphor layer for mono-energetic Al ions are shown as functions of the target depth in Fig. 5 for 130 MeV Al ions (dashed green line) and for 200 MeV Al ions (dot dashed blue line). The 50 μm thick phosphor layer of BAS-TR IP is thick enough to stop all ions with $E_{ion}$ <= 130 MeV Al ions, so that the entire $E_{ion}$ is absorbed within the layer. For 200 MeV Al ions, on the other hand, the stopping power gradually increases as the ions lose their kinetic energy, but the ions penetrate through the phosphor layer with significant energy remaining.

Figure 6 (a) shows the IP response, $R(E_{Ion})$, as a function of Al ion energy for our setup. Black dots indicate the experimental measurements of PSL/ion data. Error bars are estimated to be ±15%, which account for the decay uncertainty from the fading effect during the time gap (5 minutes), PSL stimulated by diode laser during scanning, inaccurate calibration of the PMT, and IP surface quality (roughness).[6] The dashed black line represents the IP response calculated using the linear model, and the solid red line shows the IP response calculated using the exponential model with $L$=44 μm. The calculated IP responses with ± one standard deviation (± 4 μm) of $L$ are shown when $L$=48 μm (dashed red line) and $L$=40 μm (dot dashed line). In both models, the total deposited energy increases as the kinetic energy of Al ion increases from 50 MeV to 140 MeV, but it decreases beyond 140 MeV. The IP response from the exponential model increases more gradually and it fits the experimental data better.

As mentioned earlier in the experimental setup section, there was an 18 μm thick filter in front of our BAS-TR IP, and only ions with sufficient kinetic energy (larger than 50 MeV) arrived at the IP front surface. Since filter conditions can change depending on the type of experiment, beyond the computation of our specific conditions, we aim to produce a response model applicable to arbitrary filter arrangements. We start with the computed incident ion energy on the front phosphor-layer surface obtained by subtracting the energy loss in the 18 μm thick filter from the incoming $E_{ion}$ measured with the TPS. We computed Al ion energy loss in the filter with SRIM. Figure 6 (b) shows energy loss from an 18 μm thick Al filter as a function of the incident Al ion energy. The dashed green line indicates the kinetic energy loss of energetic Al ions. The solid blue line shows the incident kinetic energy of Al ions on the phosphor layer after penetrating through the filter as a function of Al ion energy. Figure 6 (c) indicates the IP response to incident Al ion energy on the IP as if there were no filter. Our IP sensitivity and fitting coefficients of the response function are presented in Table I. All of these values were obtained by using the least-squares method.

In order to determine which model represents the IP response better, we examined how IP sensitivities at different incident Al ion energy deviate from the IP sensitivity obtained by least-square method in each model. IP sensitivities from both linear model and exponential model are plotted in Fig. 7. These values are derived from experimentally measured IP responses and the calculated deposited energy of incident Al ion energy. IP sensitivity deviation in the exponential model is less than that in the linear model. The IP sensitivity is nearly constant in the exponential model with a standard deviation of 5% of the average IP sensitivity, whereas the linear model shows a standard deviation of 20%. This shows that the exponential model describes the IP response of IP better.

The overall sensitivity of BAS-TR IP to Al ions is found to be much smaller than the known sensitivity of IPs to protons and to electrons. This is consistent with what Freeman *et al*.[22] and Bonnet *et al*.[19] found for alpha particles. Bonnet *et al*. reports that the IP sensitivity to $^4$He ions is about 10 times smaller than the IP sensitivity to protons with BAS-MS and BAS-SR IPs and around 5 times smaller than the IP sensitivity to protons with BAS-TR IP.[19] Bonnet *et al*. explains this using a quenching effect. According to their study, the IP sensitivity depends on both the type of incident ions and the stopping power of IPs for those ions.[19, 22]

In our experiment, a similar quenching effect is also observed. The measured sensitivity of BAS-TR IP to Al ions is about 13 times less than the known IP sensitivity to protons, and it is about half of the known IP sensitivity to $^4$He.[19] As shown in Fig. 7, IP sensitivities decrease with increasing incident Al ion energy. This decrease is larger in the linear model than in the exponential model. The large variations of the IP sensitivities in the linear model seem to be caused by the quenching effect, which results in large discrepancy with the experimental IP response data. In comparison, the exponential model has smaller variations of the IP sensitivity and agrees well with the experimental data. The exponential factor in Eq. (5) offsets the IP sensitivity decline coming from the stopping power increase of Al ions.

**Conclusion**

For the first time, we have measured the response of BAS-TR IP to Al ions in the 0–222 MeV range. We have calibrated BAS-TR IP using CR-39 track detectors and an ion energy spectrometer. The PSL values were measured experimentally using a commercial IP scanner, and the calibration results were compared with deposited energy. We propose a technique to calculate deposited energy in the phosphor layer using stopping power calculations from SRIM code. The response function taking the absorption length into account is in very good agreement with the experimental data. We find that the exponential model predicting the response of an IP fits the Al data better. This work paves the way towards calibration of IP with other heavy ions.

**References**

[1] Y. Amemiya, J. Miyahara, Imaging plate illuminates many fields, Nature 336 (1988) 89.
[2] Y. Amemiya, K. Wakabayashi, H. Tanaka, Y. Ueno, J. Miyahara, Laser-stimulated luminescence used to measure x-ray diffraction of a contracting striated muscle, Science 237(4811) (1987) 164.
[3] S. Cipiccia, M.R. Islam, B. Ersfeld, R.P. Shanks, E. Brunetti, G. Vieux, X. Yang, R.C. Issac, S.M. Wiggins, G.H. Welsh, M.-P. Anania, D. Maneuski, R. Montgomery, G. Smith, M. Hoek, D.J. Hamilton, N.R.C. Lemos,

D. Symes, P.P. Rajeev, V.O. Shea, J.M. Dias, D.A. Jaroszynski, Gamma-rays from harmonically resonant betatron oscillations in a plasma wake, Nature Physics 7 (2011) 867.
[4] K. Ta Phuoc, S. Corde, C. Thaury, V. Malka, A. Tafzi, J.P. Goddet, R.C. Shah, S. Sebban, A. Rousse, All-optical Compton gamma-ray source, Nature Photonics 6 (2012) 308.
[5] S. Singh, T. Slavicek, R. Hodak, R. Versaci, P. Pridal, D. Kumar, Absolute calibration of imaging plate detectors for electron kinetic energies between 150 keV and 1.75 MeV, Review of Scientific Instruments 88(7) (2017) 075105.
[6] D. Doria, S. Kar, H. Ahmed, A. Alejo, J. Fernandez, M. Cerchez, R.J. Gray, F. Hanton, D.A. MacLellan, P. McKenna, Z. Najmudin, D. Neely, L. Romagnani, J.A. Ruiz, G. Sarri, C. Scullion, M. Streeter, M. Swantusch, O. Willi, M. Zepf, M. Borghesi, Calibration of BAS-TR image plate response to high energy (3-300 MeV) carbon ions, Review of Scientific Instruments 86(12) (2015) 123302.
[7] G. Boutoux, N. Rabhi, D. Batani, A. Binet, J.E. Ducret, K. Jakubowska, J.P. Nègre, C. Reverdin, I. Thfoin, Study of imaging plate detector sensitivity to 5-18 MeV electrons, Review of Scientific Instruments 86(11) (2015) 113304.
[8] G. Boutoux, D. Batani, F. Burgy, J.E. Ducret, P. Forestier-Colleoni, S. Hulin, N. Rabhi, A. Duval, L. Lecherbourg, C. Reverdin, K. Jakubowska, C.I. Szabo, S. Bastiani-Ceccotti, F. Consoli, A. Curcio, R. De Angelis, F. Ingenito, J. Baggio, D. Raffestin, Validation of modelled imaging plates sensitivity to 1-100 keV x-rays and spatial resolution characterisation for diagnostics for the "PETawatt Aquitaine Laser", Review of Scientific Instruments 87(4) (2016) 043108.
[9] H.P. Schlenvoigt, K. Haupt, A. Debus, F. Budde, O. Jäckel, S. Pfotenhauer, H. Schwoerer, E. Rohwer, J.G. Gallacher, E. Brunetti, R.P. Shanks, S.M. Wiggins, D.A. Jaroszynski, A compact synchrotron radiation source driven by a laser-plasma wakefield accelerator, Nature Physics 4 (2007) 130.
[10] C. Yu, R. Qi, W. Wang, J. Liu, W. Li, C. Wang, Z. Zhang, J. Liu, Z. Qin, M. Fang, K. Feng, Y. Wu, Y. Tian, Y. Xu, F. Wu, Y. Leng, X. Weng, J. Wang, F. Wei, Y. Yi, Z. Song, R. Li, Z. Xu, Ultrahigh brilliance quasi-monochromatic MeV γ-rays based on self-synchronized all-optical Compton scattering, Scientific Reports 6 (2016) 29518.
[11] K. Miyasaka, T. Tanabe, G. Mank, K.H. Finken, V. Philipps, D.S. Walsh, K. Nishizawa, T. Saze, Tritium detection in plasma facing component by imaging plate technique, Journal of Nuclear Materials 290-293 (2001) 448-453.
[12] T. Otsuka, M. Shimada, R. Kolasinski, P. Calderoni, J.P. Sharpe, Y. Ueda, Y. Hatano, T. Tanabe, Application of tritium imaging plate technique to examine tritium behaviors on the surface and in the bulk of plasma-exposed materials, Journal of Nuclear Materials 415(1, Supplement) (2011) S769-S772.
[13] K. Sugiyama, T. Tanabe, K. Masaki, N. Miya, Tritium distribution measurement of the tile gap of JT-60U, Journal of Nuclear Materials 367-370 (2007) 1248-1253.
[14] F. Ingenito, P. Andreoli, D. Batani, G. Boutoux, M. Cipriani, F. Consoli, G. Cristofari, A. Curcio, R.D. Angelis, G.D. Giorgio, J. Ducret, P. Forestier-Colleoni, S. Hulin, K. Jakubowska, N. Rabhi, Comparative calibration of IP scanning equipment, Journal of Instrumentation 11(05) (2016) C05012.
[15] N. Rabhi, D. Batani, G. Boutoux, J.-E. Ducret, K. Jakubowska, I. Lantuejoul-Thfoin, C. Nauraye, A. Patriarca, A. Saïd, A. Semsoum, L. Serani, B. Thomas, B. Vauzour, Calibration of imaging plate detectors to mono-energetic protons in the range 1-200 MeV, Review of Scientific Instruments 88(11) (2017) 113301.
[16] K. Zeil, S.D. Kraft, A. Jochmann, F. Kroll, W. Jahr, U. Schramm, L. Karsch, J. Pawelke, B. Hidding, G. Pretzler, Absolute response of Fuji imaging plate detectors to picosecond-electron bunches, Review of Scientific Instruments 81(1) (2010) 013307.
[17] T. Bonnet, M. Comet, D. Denis-Petit, F. Gobet, F. Hannachi, M. Tarisien, M. Versteegen, M.M. Aleonard, Response functions of Fuji imaging plates to monoenergetic protons in the energy range 0.6–3.2 MeV, Review of Scientific Instruments 84(1) (2013) 013508.

[18] B.R. Maddox, H.S. Park, B.A. Remington, N. Izumi, S. Chen, C. Chen, G. Kimminau, Z. Ali, M.J. Haugh, Q. Ma, High-energy x-ray backlighter spectrum measurements using calibrated image plates, Review of Scientific Instruments 82(2) (2011) 023111.
[19] T. Bonnet, M. Comet, D. Denis-Petit, F. Gobet, F. Hannachi, M. Tarisien, M. Versteegen, M.M. Aléonard, Response functions of imaging plates to photons, electrons and 4He particles, Review of Scientific Instruments 84(10) (2013) 103510.
[20] A. Mančić, J. Fuchs, P. Antici, S.A. Gaillard, P. Audebert, Absolute calibration of photostimulable image plate detectors used as (0.5–20MeV) high-energy proton detectors, Review of Scientific Instruments 79(7) (2008) 073301.
[21] I.W. Choi, C.M. Kim, J.H. Sung, I.J. Kim, T.J. Yu, S.K. Lee, Y.Y. Jin, K.H. Pae, N. Hafz, J. Lee, Absolute calibration of a time-of-flight spectrometer and imaging plate for the characterization of laser-accelerated protons, Measurement Science and Technology 20(11) (2009) 115112.
[22] C.G. Freeman, G. Fiksel, C. Stoeckl, N. Sinenian, M.J. Canfield, G.B. Graeper, A.T. Lombardo, C.R. Stillman, S.J. Padalino, C. Mileham, T.C. Sangster, J.A. Frenje, Calibration of a Thomson parabola ion spectrometer and Fujifilm imaging plate detectors for protons, deuterons, and alpha particles, Review of Scientific Instruments 82(7) (2011) 073301.
[23] K.A. Tanaka, T. Yabuuchi, T. Sato, R. Kodama, Y. Kitagawa, T. Takahashi, T. Ikeda, Y. Honda, S. Okuda, Calibration of imaging plate for high energy electron spectrometer, Review of Scientific Instruments 76(1) (2005) 013507.
[24] H. Chen, N.L. Back, T. Bartal, F.N. Beg, D.C. Eder, A.J. Link, A.G. MacPhee, Y. Ping, P.M. Song, A. Throop, L. Van Woerkom, Absolute calibration of image plates for electrons at energy between 100keV and 4MeV, Review of Scientific Instruments 79(3) (2008) 033301.
[25] N. Nakanii, K. Kondo, T. Yabuuchi, K. Tsuji, K.A. Tanaka, S. Suzuki, T. Asaka, K. Yanagida, H. Hanaki, T. Kobayashi, K. Makino, T. Yamane, S. Miyamoto, K. Horikawa, Absolute calibration of imaging plate for GeV electrons, Review of Scientific Instruments 79(6) (2008) 066102.
[26] A. Alejo, S. Kar, H. Ahmed, A.G. Krygier, D. Doria, R. Clarke, J. Fernandez, R.R. Freeman, J. Fuchs, A. Green, J.S. Green, D. Jung, A. Kleinschmidt, C.L.S. Lewis, J.T. Morrison, Z. Najmudin, H. Nakamura, G. Nersisyan, P. Norreys, M. Notley, M. Oliver, M. Roth, J.A. Ruiz, L. Vassura, M. Zepf, M. Borghesi, Characterisation of deuterium spectra from laser driven multi-species sources by employing differentially filtered image plate detectors in Thomson spectrometers, Review of Scientific Instruments 85(9) (2014) 093303.
[27] B. Hidding, G. Pretzler, M. Clever, F. Brandl, F. Zamponi, A. Lübcke, T. Kämpfer, I. Uschmann, E. Förster, U. Schramm, R. Sauerbrey, E. Kroupp, L. Veisz, K. Schmid, S. Benavides, S. Karsch, Novel method for characterizing relativistic electron beams in a harsh laser-plasma environment, Review of Scientific Instruments 78(8) (2007) 083301.
[28] T. Bonnet, M. Comet, D. Denis-Petit, F. Gobet, F. Hannachi, M. Tarisien, M. Versteegen, Two parameter model of Fuji imaging plate response function to protons, SPIE Optics + Optoelectronics, SPIE, 2013, p. 7.
[29] N. Rabhi, K. Bohacek, D. Batani, G. Boutoux, J.E. Ducret, E. Guillaume, K. Jakubowska, C. Thaury, I. Thfoin, Calibration of imaging plates to electrons between 40 and 180 MeV, Review of Scientific Instruments 87(5) (2016) 053306.
[30] J.F. Ziegler, M.D. Ziegler, J.P. Biersack, SRIM – The stopping and range of ions in matter (2010), Nuclear Instruments and Methods in Physics Research Section B: Beam Interactions with Materials and Atoms 268(11) (2010) 1818-1823.
[31] H. Paul, D. Sánchez-Parcerisa, A critical overview of recent stopping power programs for positive ions in solid elements, Nuclear Instruments and Methods in Physics Research Section B: Beam Interactions with Materials and Atoms 312 (2013) 110-117.

[32] I.G. Evseev, H.R. Schelin, S.A. Paschuk, E. Milhoretto, J.A.P. Setti, O. Yevseyeva, J.T. de Assis, J.M. Hormaza, K.S. Díaz, R.T. Lopes, Comparison of SRIM, MCNPX and GEANT simulations with experimental data for thick Al absorbers, Applied Radiation and Isotopes 68(4) (2010) 948-950.
[33] S. Palaniyappan, C. Huang, D.C. Gautier, C.E. Hamilton, M.A. Santiago, C. Kreuzer, A.B. Sefkow, R.C. Shah, J.C. Fernández, Efficient quasi-monoenergetic ion beams from laser-driven relativistic plasmas, Nature Communications 6 (2015) 10170.
[34] W. Bang, B.J. Albright, P.A. Bradley, D.C. Gautier, S. Palaniyappan, E.L. Vold, M.A.S. Cordoba, C.E. Hamilton, J.C. Fernández, Visualization of expanding warm dense gold and diamond heated rapidly by laser-generated ion beams, Scientific Reports 5 (2015) 14318.
[35] W. Bang, B.J. Albright, P.A. Bradley, E.L. Vold, J.C. Boettger, J.C. Fernández, Linear dependence of surface expansion speed on initial plasma temperature in warm dense matter, Scientific Reports 6 (2016) 29441.
[36] W. Bang, B.J. Albright, P.A. Bradley, E.L. Vold, J.C. Boettger, J.C. Fernández, Uniform heating of materials into the warm dense matter regime with laser-driven quasimonoenergetic ion beams, Physical Review E 92(6) (2015) 063101.
[37] J.C. Fernández, D. Cort Gautier, C. Huang, S. Palaniyappan, B.J. Albright, W. Bang, G. Dyer, A. Favalli, J.F. Hunter, J. Mendez, M. Roth, M. Swinhoe, P.A. Bradley, O. Deppert, M. Espy, K. Falk, N. Guler, C. Hamilton, B.M. Hegelich, D. Henzlova, K.D. Ianakiev, M. Iliev, R.P. Johnson, A. Kleinschmidt, A.S. Losko, E. McCary, M. Mocko, R.O. Nelson, R. Roycroft, M.A. Santiago Cordoba, V.A. Schanz, G. Schaumann, D.W. Schmidt, A. Sefkow, T. Shimada, T.N. Taddeucci, A. Tebartz, S.C. Vogel, E. Vold, G.A. Wurden, L. Yin, Laser-plasmas in the relativistic-transparency regime: Science and applications, Physics of Plasmas 24(5) (2017) 056702.
[38] D. Jung, R. Hörlein, D. Kiefer, S. Letzring, D.C. Gautier, U. Schramm, C. Hübsch, R. Öhm, B.J. Albright, J.C. Fernandez, D. Habs, B.M. Hegelich, Development of a high resolution and high dispersion Thomson parabola, Review of Scientific Instruments 82(1) (2011) 013306.
[39] G. Schiwietz, P.L. Grande, Improved charge-state formulas, Nuclear Instruments and Methods in Physics Research Section B: Beam Interactions with Materials and Atoms 175-177 (2001) 125-131.
[40] H.-D. Betz, Charge States and Charge-Changing Cross Sections of Fast Heavy Ions Penetrating Through Gaseous and Solid Media, Reviews of Modern Physics 44(3) (1972) 465-539.
[41] H. Ohuchi, A. Yamadera, Dependence of fading patterns of photo-stimulated luminescence from imaging plates on radiation, energy, and image reader, Nuclear Instruments and Methods in Physics Research Section A: Accelerators, Spectrometers, Detectors and Associated Equipment 490(3) (2002) 573-582.
[42] See http://srim.org/.

**Acknowledgements**

This work was supported by NRF-2018R1C1B6001580. The experimental work was performed at LANL, operated by Los Alamos National Security, LLC, for the U.S. DOE under Contract No. DE-AC52-06NA25396, and was supported in part by the LANL LDRD program. J.W., J.S., and W.B. were supported by Institute for Basic Science under IBS-R012-D1-2018-A00. The authors would like to thank M. Cordoba and C.E. Hamilton for target fabrication, and thank the Trident laser team for operation of the laser and assistance with the implementation of TPS diagnostics.

**Table and Figures**

| Fitting Function | | $y = \alpha(A_1 x + A_3 x^3 + A_5 x^5 + A_7 x^7)$ | | | $y = \alpha(B x^C)$ | |
|---|---|---|---|---|---|---|
| | | 0 – 104 MeV | Standard error | | 104 – 222 MeV | Standard error |
| **Exponential model (L=44 μm)** | $A_1$ | $8.89153 \times 10^{-1}$ | $\pm 1.649 \times 10^{-3}$ | B | 1414.56 | $\pm 11.37$ |
| | $A_3$ | $6.07554 \times 10^{-5}$ | $\pm 1.1106 \times 10^{-6}$ | C | - 0.662522 | $\pm 9.837 \times 10^{-4}$ |
| | $A_5$ | $4.94672 \times 10^{-9}$ | $\pm 2.138 \times 10^{-10}$ | α (PSL/MeV) | | 0.019024 |
| | $A_7$ | $2.21614 \times 10^{-13}$ | $\pm 1.219 \times 10^{-14}$ | | | |

**TABLE I.** Coefficients of PSL/ion fitting functions and IP sensitivities for the exponential model.

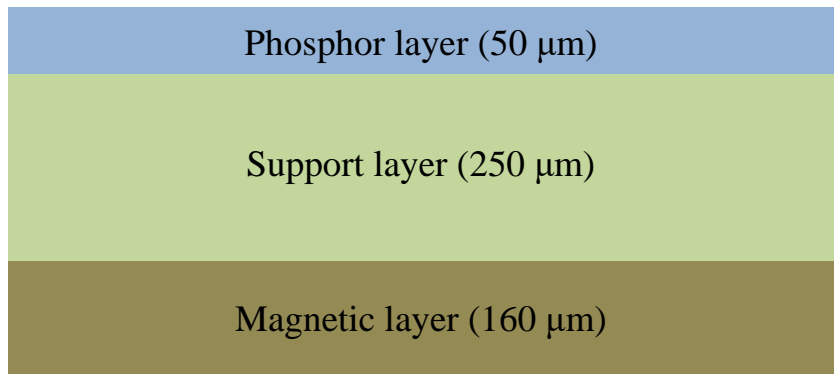

**FIG. 1.** Structure of BAS-TR imaging plate.[14]

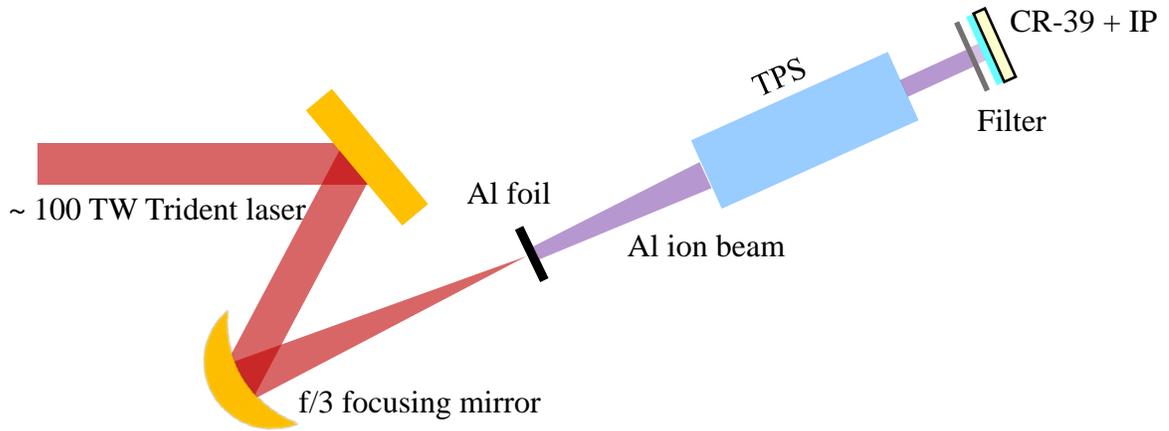

**FIG. 2.** Schematic layout of the experimental setup. An intense laser pulse produces energetic Al ions, which are detected using CR-39 and IP.

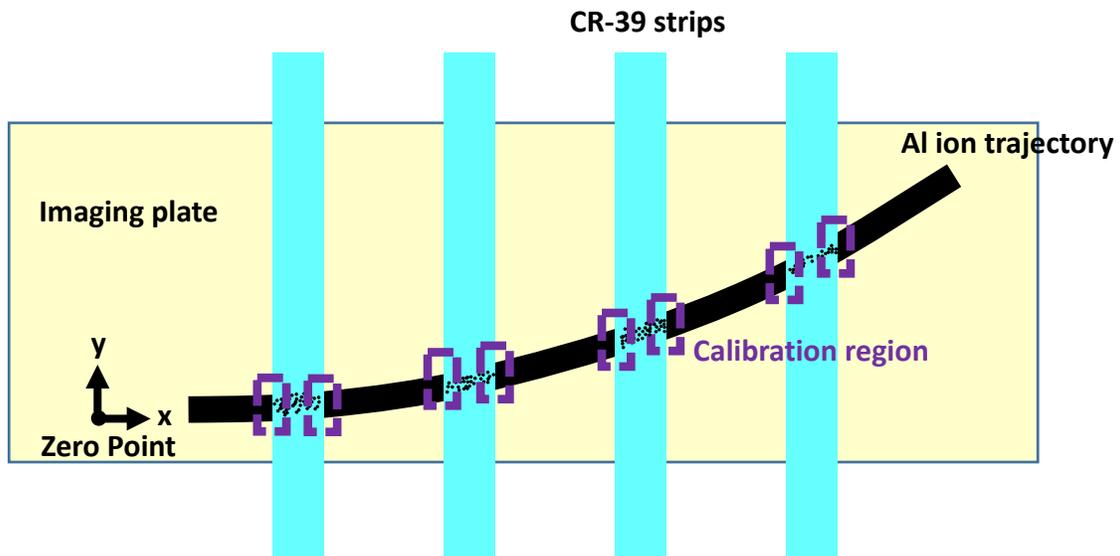

**FIG. 3.** Setup of CR-39 strips and a BAS-TR IP. The trajectory of Al ions is formed on both CR-39 strips and BAS-TR IP. At each boundary, PSL/ion can be measured from the number of pits on the CR-39 strip and the amount of light emitted from the IP.

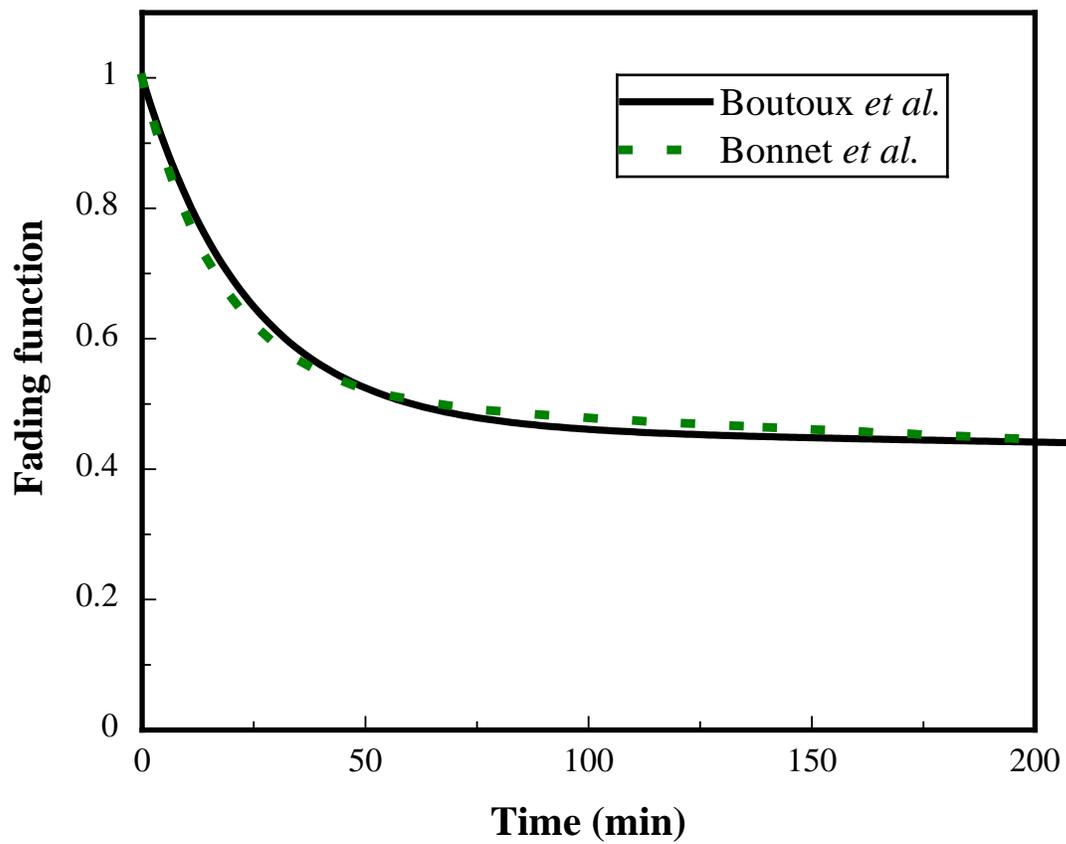

**FIG. 4.** Comparison of fading function between Boutoux *et al.* and Bonnet *et al.*

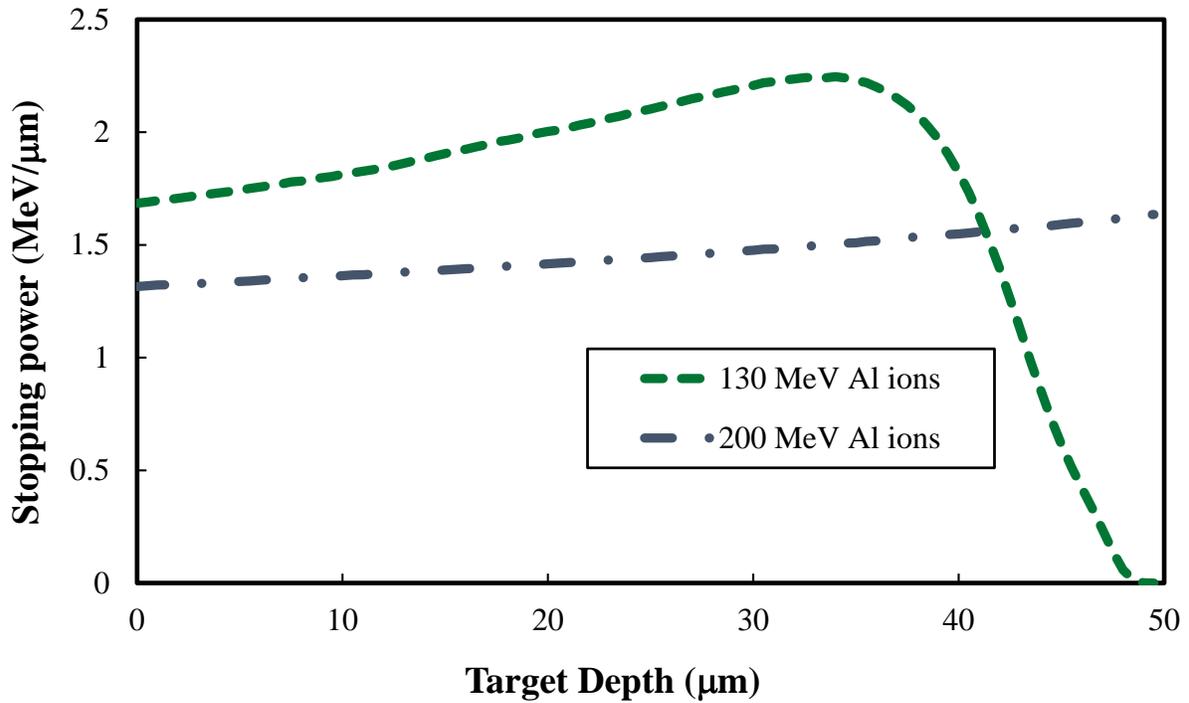

**FIG. 5.** Stopping power of the 50 μm thick phosphor layer of a BAS-TR IP for 130 MeV and 200 MeV Al ions as functions of the target depth. The area under the curve represents the deposited energy in the phosphor layer.

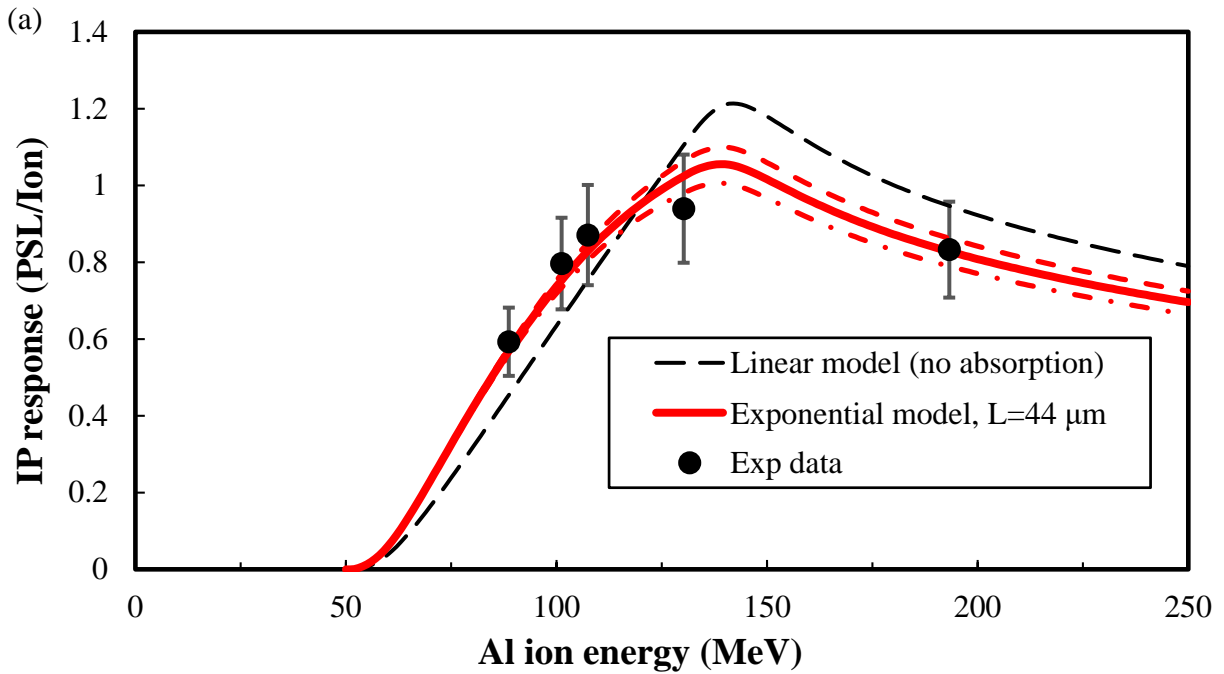

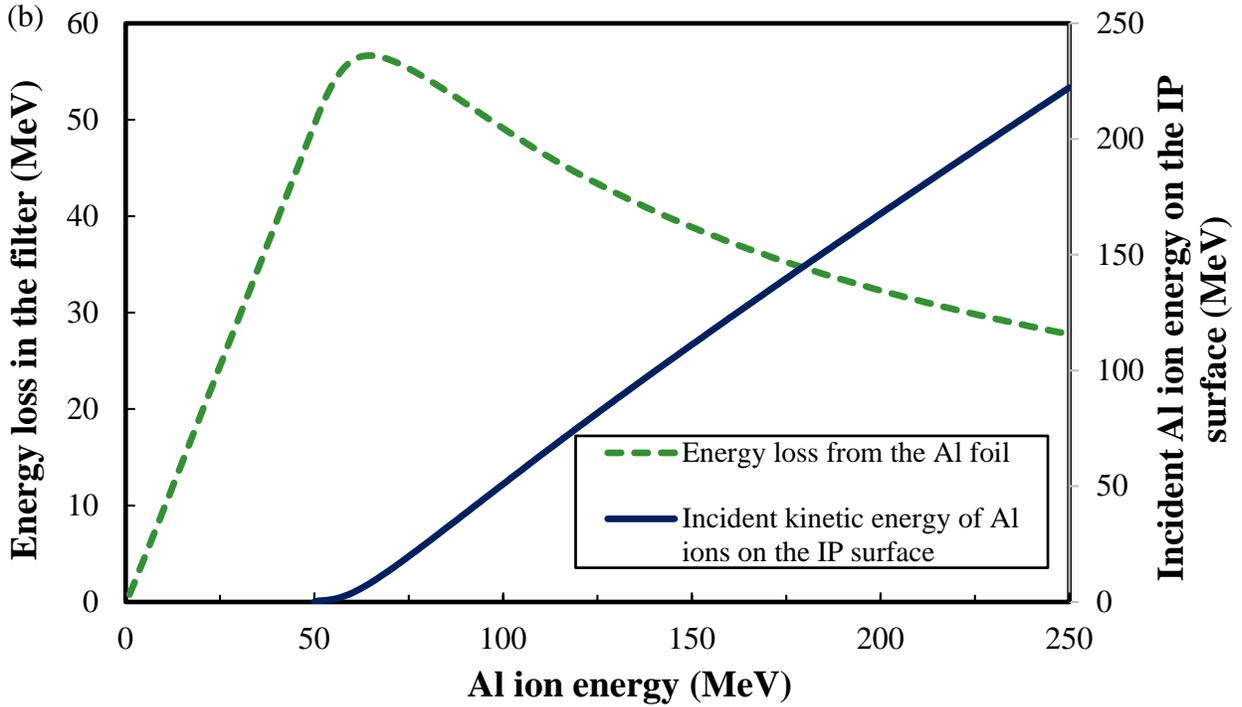

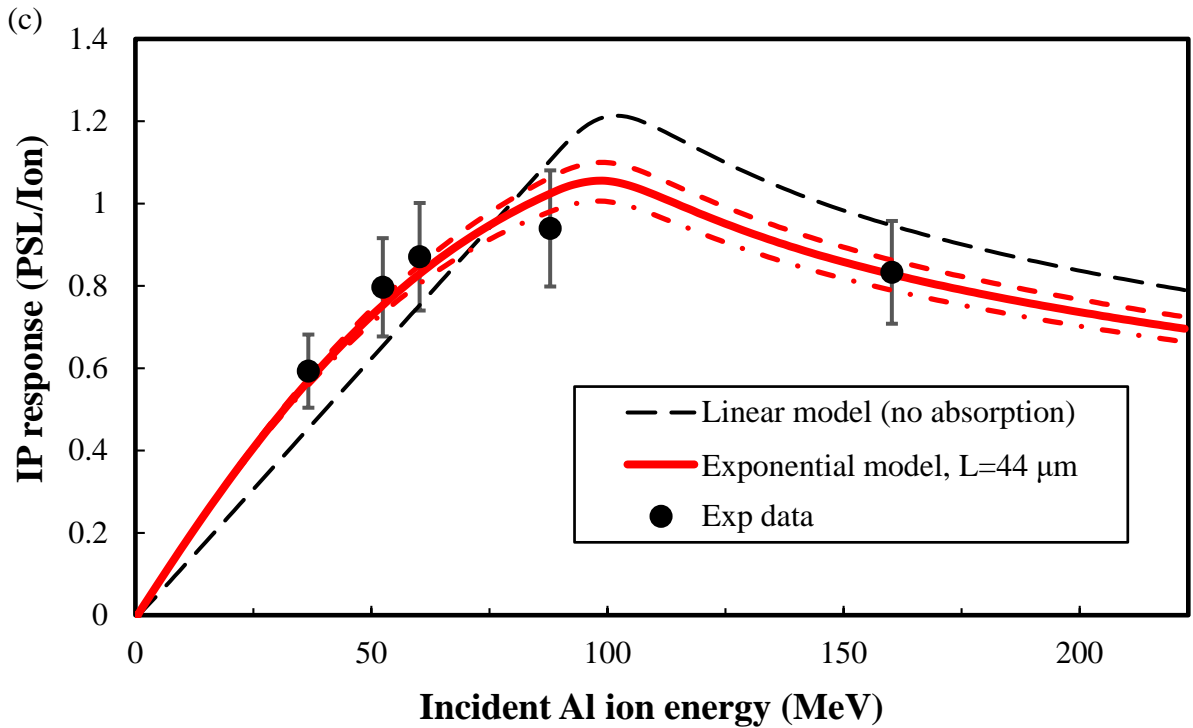

**FIG. 6.** (a) IP response of BAS-TR IP is shown as a function of Al ion energy. Black dots indicate the experimental measurements of PSL/ion data. Solid red line shows the IP response calculated using the exponential model with $L$=44 μm. The IP responses with ± one standard

deviation of absorption length are expressed when *L*=48 μm (dashed red line) and *L*=40 μm (dot dashed line). (b) Energy loss in the filter is shown as a function of Al ion energy (dashed green line). The incident kinetic energy of Al ions on the phosphor layer after penetrating through the filter is also shown as a function of Al ion energy (solid blue line). (c) IP response of BAS-TR IP is shown as a function of the incident Al ion energy on the phosphor layer. Solid red line shows the IP response calculated using the exponential model with *L*=44 μm. The IP responses with ± one standard deviation of absorption length are expressed when *L*=48 μm (dashed red line) and *L*=40 μm (dot dashed line).

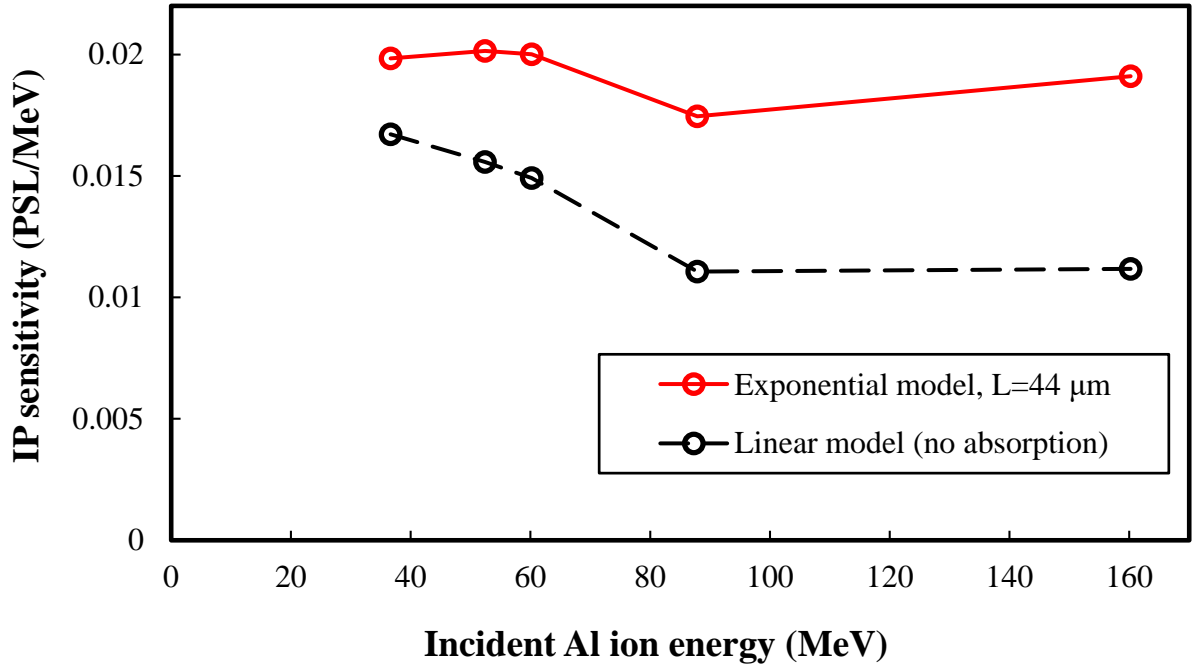

**FIG. 7.** Sensitivity of BAS-TR IP is shown as a function of the incident Al ion energy on the IP surface.